\documentclass[12pt]{iopart}

\usepackage{graphics}

\begin{document}

\title[]{Design, Fabrication, and Experimental Demonstration of Junction Surface Ion Traps}

\author{D.~L.~Moehring, C.~Highstrete, D.~Stick, K.~M.~Fortier, R.~Haltli, C.~Tigges, and M.~G.~Blain}

\address{Sandia National Laboratories, Albuquerque, New Mexico 87185, USA}
\ead{dlmoehr@sandia.gov}

\begin{abstract}
We present the design, fabrication, and experimental implementation of surface ion traps with Y-shaped junctions.  The traps are designed to minimize the pseudopotential variations in the junction region at the symmetric intersection of three linear segments.  We experimentally demonstrate robust linear and junction shuttling with greater than $10^6$ round-trip shuttles without ion loss.  By minimizing the direct line of sight between trapped ions and dielectric surfaces, negligible day-to-day and trap-to-trap variations are observed.  In addition to high-fidelity single-ion shuttling, multiple-ion chains survive splitting, ion-position swapping, and recombining routines.  The development of two-dimensional trapping structures is an important milestone for ion-trap quantum computing and quantum simulations.  
\end{abstract}


\maketitle

\section{Introduction}
\label{sec:intro}
The first requirement for quantum information processing is the ability to build a ``scalable physical system with well characterized qubits'' \cite{divincenzo:2000}.  In recent years, research in trapped ion quantum information has concentrated much effort toward creating scalable architectures for trapping and shuttling large numbers of ions \cite{wineland:1998,kielpinski:2002,rowe:2002,madsen:2004,stick:2006,hensinger:2006,reichle:2006,brownnut:2006,leibfried:2007,amini:2008,blakestad:2009,huges:2011}.  Within this effort, surface-electrode ion traps are generally regarded as the most promising long-term approach due to the ability to fabricate complex trap arrays leveraging advanced semiconductor and microfabrication techniques \cite{chiaverini:2005,britton:2006,seidelin:2007,brown:2007,britton:2009,leibrandt:2009,allcock:2010,hellwig:2010,amini:2010,stick:2010,ospelkaus:2011}.  Two-dimensional ion-trap geometries, such as those shown here, and ion trap arrays are key developments for scalable implementations of ion-based quantum computation and simulation \cite{schmied:2009,barreiro:2011,kumph:2011}.  

In this paper, we report the design, fabrication, and successful testing of Y-junction surface ion traps.  We demonstrate high-fidelity shuttling protocols in three different traps of two different trap designs.  This includes linear and junction shuttling, splitting and recombining of ion chains, and ion reordering.  The surface microtraps reported here are both reproducible and invariable.  This is demonstrated by successful ion-shuttling solutions that are identical for day-to-day operation as well as for multiple congeneric traps.

\section{Design and Fabrication}
\label{sec:design}
\begin{figure}
\resizebox{1\textwidth}{!}{%
  \includegraphics{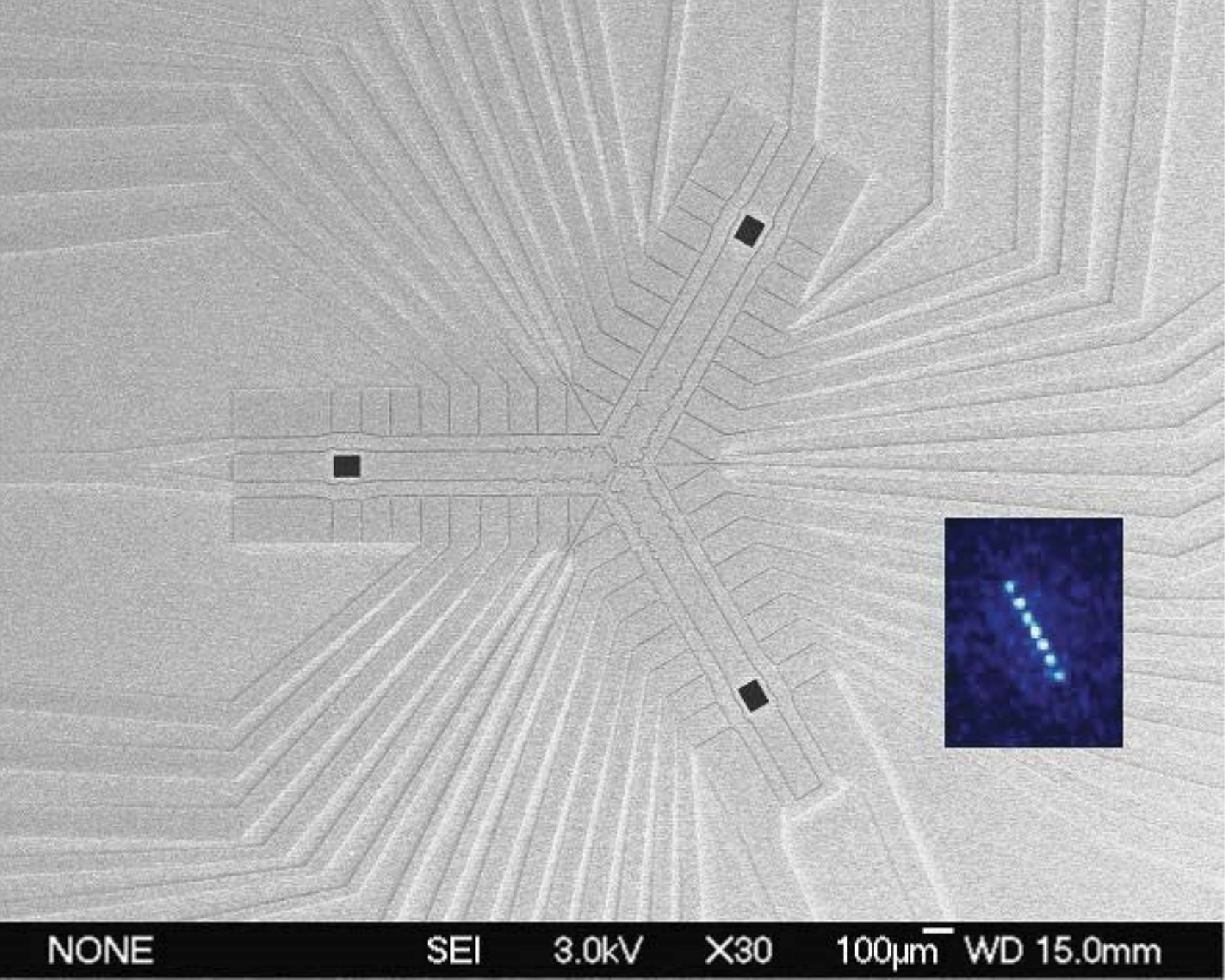}}
\caption{Scanning electron microscope (SEM) image of a Y$_{\textrm{H}}$-junction trap.  In each trap, 47 independent DC electrodes are routed for wire bonding to the CPGA.  Inset: Image of 7 ions trapped above the loading hole.  The average ion-ion spacing in this image is $\approx3.5~\mu$m}
\label{fig:trap}       
\end{figure}

The design principles for the traps discussed here are based in part on fabrication constraints and techniques described in our previous work \cite{stick:2010}.  These principles include minimized line of sight exposure between the ion and dielectric surfaces to ensure shielding of any trapped dielectric charge \cite{harlander:2010}, and recessed bond pads to limit the projection of the wire-bond ribbons above the top surface of the trap.  In addition, the top metal layout in the vicinity of the junction and the loading hole is optimized with respect to RF potential and other characteristics by utilizing lateral shape-modulation of electrode edges [Fig.~\ref{fig:trap}].

\begin{figure}
\resizebox{1\textwidth}{!}{%
  \includegraphics{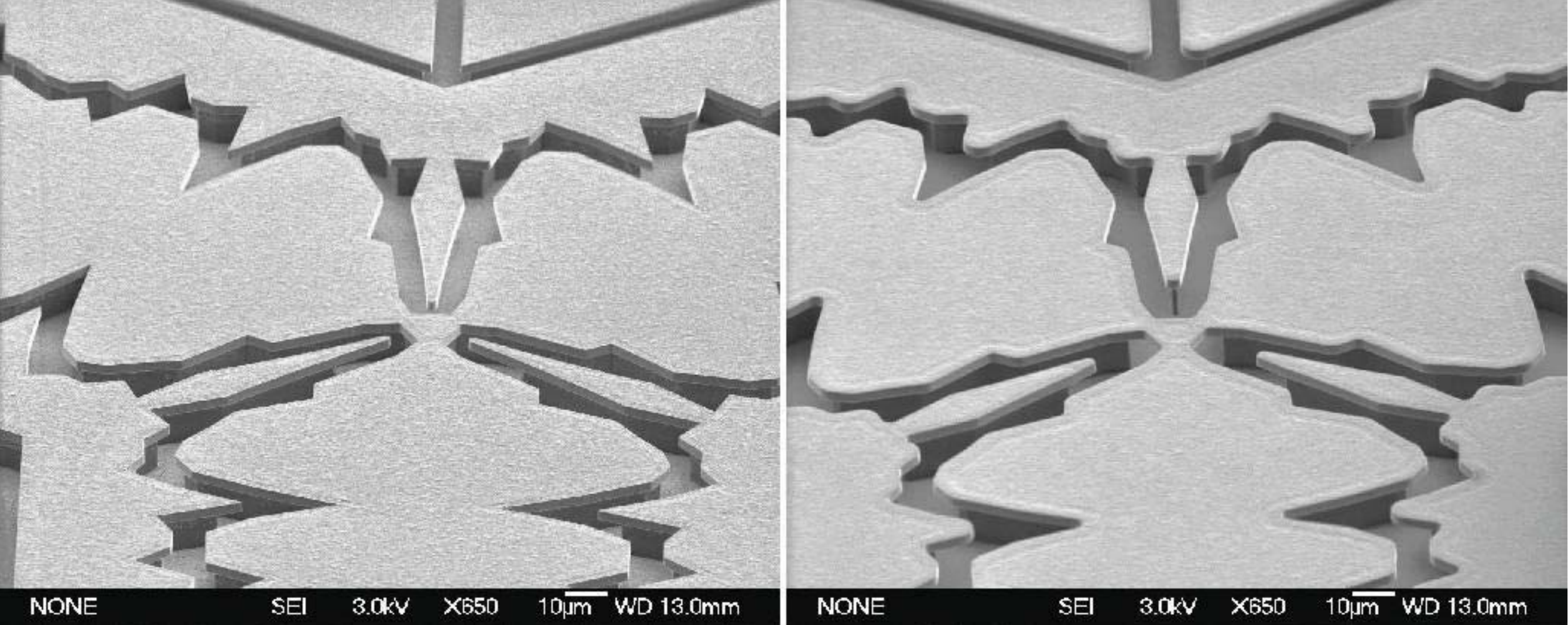}}
\caption{SEM image of the central region of the Y$_{\textrm{H}}$-junction trap (left) and the Y$_{\textrm{L}}$-junction trap (right).  Note especially the sharp edges in Y$_{\textrm{H}}$ that are rounded in Y$_{\textrm{L}}$.}
\label{fig:SEM_Junction}       
\end{figure}

\begin{figure}
\resizebox{1\textwidth}{!}{%
  \includegraphics{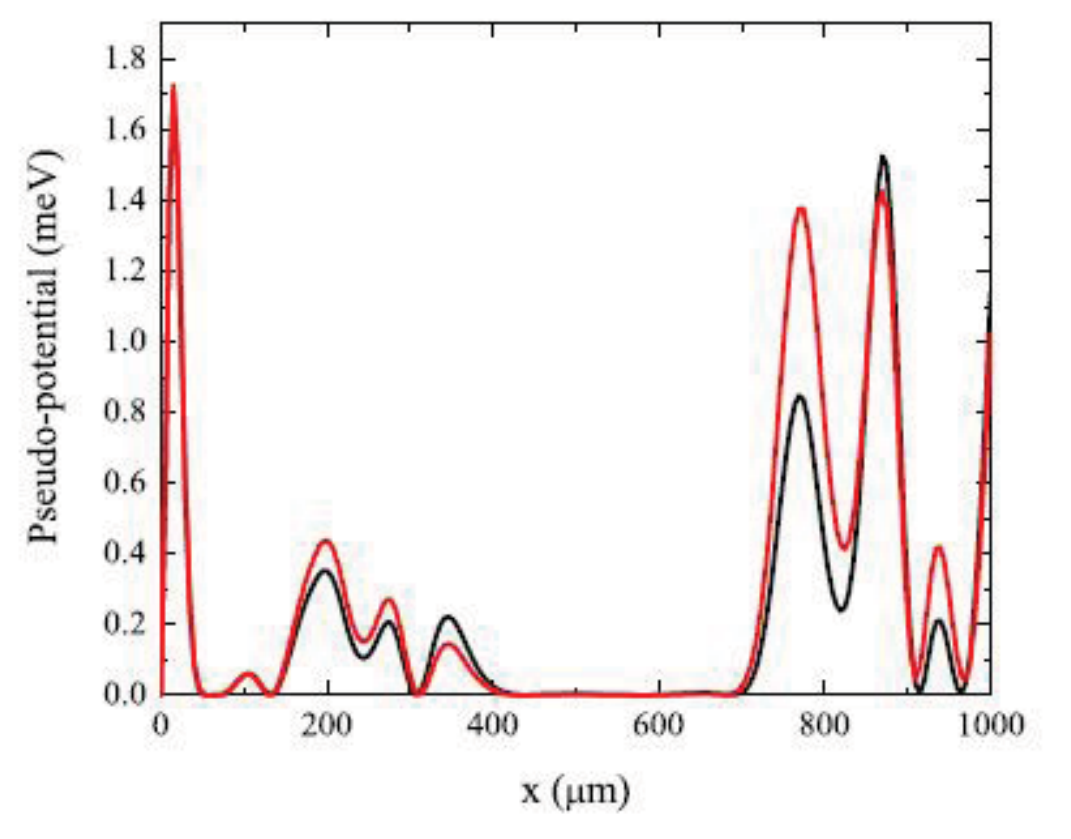}}
\caption{Two different BE simulations of the trap pseudopotential.  Both show similar results from the junction center ($x = 0~\mu$m) outward along a linear segment of the Y$_{\textrm{H}}$ trap.  The loading hole is centered at $x = 853~\mu$m.}
\label{fig:psuedopotential}       
\end{figure}

As with all known junction traps, it is not possible to stabilize a trapped charge with a vanishing RF field and zero ponderomotive potential everywhere \cite{wesenberg:2009}.  However, a suitable trade in performance characteristics can be achieved by optimizing the predicted performance using a figure of merit as a function of the electrode geometry.  The trap geometries chosen here placed particular emphasis on minimizing the magnitude and slope of the equilibrium pseudopotential particularly in the junction region while maintaining a specified ion equilibrium height range.  Resulting junction designs are shown in Figures~\ref{fig:trap} and \ref{fig:SEM_Junction}. The spatial features of the RF electrodes in the junction region decrease the pseudopotential barrier from greater than 1~eV for straight RF electrodes to less than 2~meV so that the ion can be transported through the junction with relatively reduced control voltages and motional heating~\cite{blakestad:2009} [Fig.~\ref{fig:psuedopotential}].

\begin{figure}
\resizebox{1\textwidth}{!}{%
  \includegraphics{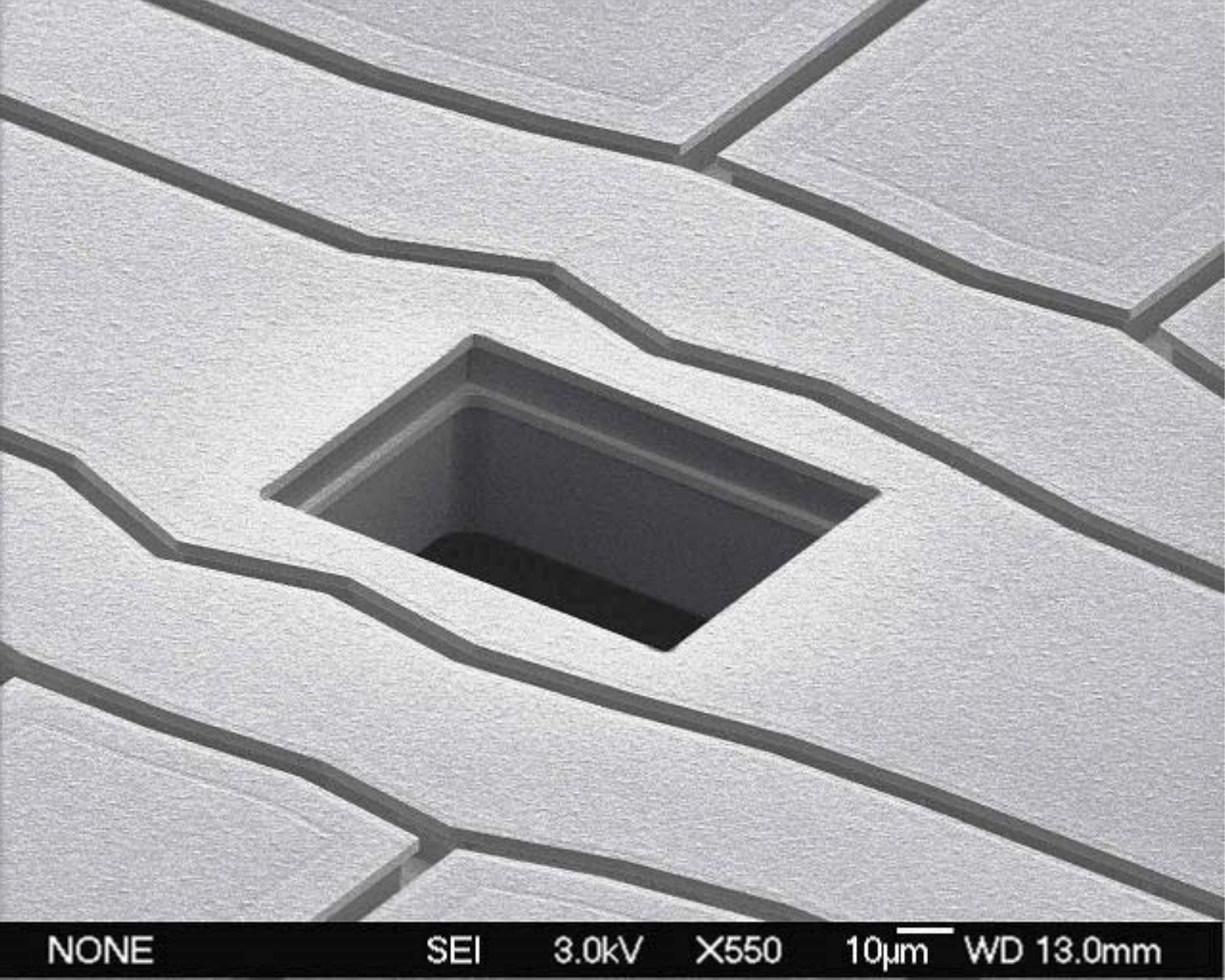}}
\caption{SEM image showing the detail of the loading hole and the modulated edges of nearby electrodes.}
\label{fig:SEM_Loading}       
\end{figure}

Another feature of these traps is the (70~$\mu$m $\times$ 86~$\mu$m) loading hole [Fig.~\ref{fig:SEM_Loading}].  The presence of the loading hole, in contrast to a solid center DC electrode, also gives rise to pseudopotential barriers that can be reduced by modulating the edges of nearby RF and DC electrodes.  The distance between the center of the loading hole and the center of the junction is $853~\mu$m.

Electrostatic solutions were calculated using a boundary element (BE) method to predict device performance.  The BE models were created by describing electrode geometries in the form of planar polygons in three spatial dimensions in conjunction with additional meta information (i.e. ion equilibrium position, BE mesh length scale) to control model fidelity.  In these models, the dielectric contribution has been ignored since these materials are always well shielded by metal.  The geometries were determined by a parametric description in order to facilitate computer optimization.

Optimization of the electrode layout was accomplished by minimizing a design cost function described by parametric values for a specific geometry.  This cost function was a sum of several positive semi-definite sub-cost terms that were traded by weighting coefficients.  The cost functions included figures of merit for the ion height and the pseudopotential values and derivatives along the equilibrium trap axis of one arm of the Y-junction.  The sub-cost values at points along the trap axis were not always treated uniformly -- some points in particular were weighted differently to effect a desired outcome not achieved by solely varying the trade coefficients.  For example, the ion height near the junction was reduced in cost so as to not trade so heavily against the pseudopotential minimization.

Two different trap models were fabricated and tested.  Trap ``Y$_{\textrm{L}}$'' has lower spatial frequencies on the electrode edge shapes in the junction region, whereas the second trap ``Y$_{\textrm{H}}$'' has higher spatial frequencies [Fig.~\ref{fig:SEM_Junction}].  The higher spatial frequencies in Y$_{\textrm{H}}$ are predicted to further reduce the pseudopotential by $\approx50\%$ compared to Y$_{\textrm{L}}$.  Two trap versions were fabricated due to a concern over the structural integrity of the cantilevered electrode segments, caused by dielectric setbacks in the high spatial-frequency regions of Y$_{\textrm{H}}$ \cite{stick:2010}.  This concern was also addressed by using a dielectric setback of only 2~$\mu$m for Y$_{\textrm{H}}$, as opposed to 5~$\mu$m for Y$_{\textrm{L}}$.  In the end, the structural integrity of the aluminum was not an issue.

\section{Operation}
\label{sec:operation}

Each fabricated ion-trap chip is packaged in a 100-pin ceramic pin grid array (CPGA) providing electrical connections to the RF electrode and 47 independent DC electrodes.  Each trap is installed in a vacuum chamber with a base pressure $\approx5\times10^{-11}$~Torr.  The applied DC voltages are varied between -10~V and +10~V using National Instruments PXI-6733 DAC cards.  The applied RF voltage is varied between 25 - 165 V for trapping in the loading hole and 85 - 120 V for junction shuttling.  Trapping lifetimes are several hours when the ions are Doppler cooled, and approximately one minute without laser cooling.

Calcium ions are loaded by first generating a stream of neutral calcium atoms through the loading hole [Fig.~\ref{fig:SEM_Loading}], which are then photoionized using a resonant 423~nm laser on the 4s$^1$S$_0\leftrightarrow$~4p$^1$P$_1$ transition and an ionizing 375~nm laser \cite{gulde:2001}.  By placing the source of the atomic beam beneath the chip, we minimize neutral atom plating on the top surface of the chip, virtually eliminating any chance of shorting adjacent trap electrodes.  This is essential for consistent day-to-day trap operations \cite{daniilidis:2011}.   

\begin{figure}
\resizebox{0.65\textwidth}{!}{%
  \includegraphics{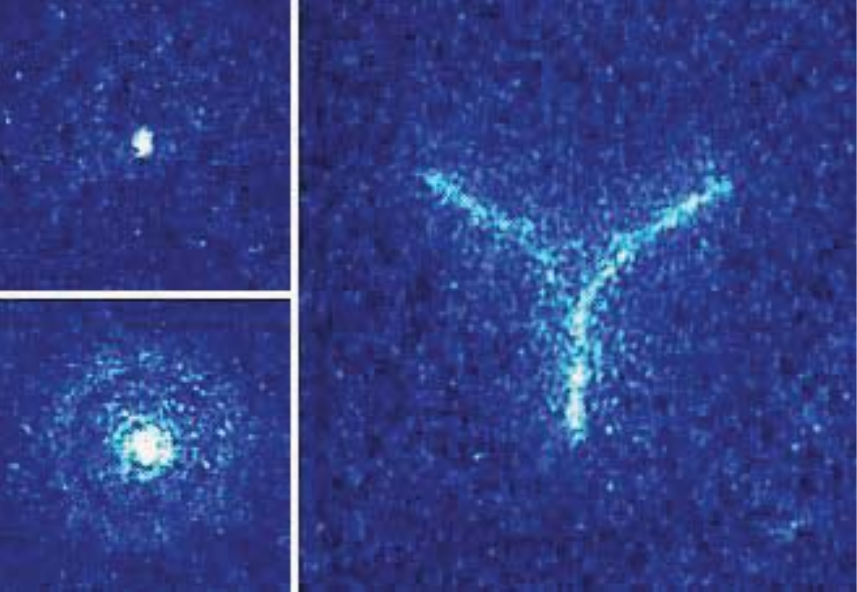}}
\caption{Left: Direct image of the ion (upper) and indirect image of the ion reflected by the aluminum trap surface (lower).  Right: Elapsed-time image of an ion shuttling $\approx40$ microns up each arm in a Y$_{\textrm{H}}$ trap.  This 10 second image captures $\approx40,000$ round-trip junction shuttles.  Rather than shuttling into the exact center of the junction, the ions are intentionally steered along a smooth path between the linear sections.}
\label{fig:height_Y}       
\end{figure}

The height of the ions above the top trap surface can be directly measured after shuttling the ions from the loading hole.  This is accomplished by imaging the ion directly and imaging the reflected photons off the aluminum center trap electrode [Figure~\ref{fig:height_Y}].  The translation of the imaging lens between the two images confirms the expected $\approx70~\mu$m height of the ion above the trap surface \cite{allcock:2010,herskind:2011}.

In total, one Y$_{\textrm{L}}$ trap and two Y$_{\textrm{H}}$ traps [Fig.~\ref{fig:SEM_Junction}] were tested, with initial work performed on the Y$_{\textrm{L}}$ trap.  Here, trapping and linear shuttling tests were completed as in \cite{stick:2010} with greater than $10^5$ linear shuttles without ion loss.  Junction shuttling was also tested in the Y$_{\textrm{L}}$ trap, and despite the lower spatial frequency electrode variations, we performed $10^6$ round-trip shuttles without ion loss.  In a given round-trip, the ion traveled up each leg of the Y-junction by $\approx30~\mu$m, resulting in a total of $3\times10^6$ passes through the junction.  

Following these initial successes, testing began in two congeneric Y$_{\textrm{H}}$ traps in two independent test systems.  As the Y$_{\textrm{H}}$ trap design differs from Y$_{\textrm{L}}$ only in the junction region, successful loading and linear shuttling voltage solutions used for Y$_{\textrm{L}}$ were demonstrated to work also in each Y$_{\textrm{H}}$ trap.  In contrast, voltage solutions for ion transfer through the modified junction electrodes required a new shuttling routine.  Interestingly, successful junction shuttling was observed in each Y$_{\textrm{H}}$ trap with a voltage-solution modification consisting of a -0.5 volt adjustment on only the center electrode \cite{footnote:centerelectrode,footnote:reversecompatibility}.  

Multiple ions were also shuttled with these solutions from the loading hole, thrice through the junction, and back to the loading hole.  One round trip reverses the order of ions within the linear chain, however this is difficult to prove unequivocally as all ions were the same isotope ($^{40}$Ca$^{+}$).  All of the above mentioned tests utilized high-degree-of-freedom voltage solutions which used up to 25 DC electrodes at a given time for linear shuttling and 35 DC electrodes for junction shuttling.

After success with the high-degree-of-freedom solutions, we tested voltage solutions with a reduced number of DC electrodes.  This is convenient for parallel shuttling of ions in multiple harmonic wells with minimal crosstalk.  It is also convenient to use voltage shuttling solutions with a constant center DC electrode \cite{footnote:centerelectrode}.  With solutions utilizing only the nearest 7 DC electrodes at any one time in the linear regions and the central 13 DC electrodes in the junction region (including the center electrode), ions were shuttled to all functional regions of each trap.  Ions survive these routines with or without Doppler cooling during shuttling.  

As in the Y$_{\textrm{L}}$ trap, a single ion in a Y$_{\textrm{H}}$ trap successfully completed $10^6$ round-trip shuttles around the junction without ion loss.  With the ion moving $40$ microns up each arm, $10^6$ shuttles took about 4 minutes, whereas moving $250$ microns up each arm took about 24 minutes.  In the latter case, the ion traveled a total distance of $1.5$~km at an average speed of $1~$m/s.  Utilizing Doppler cooling during the junction shuttling routine, the emission of photons from the ion resulted in the trace of the ion path shown in Figure~\ref{fig:height_Y}.  Importantly, voltage solutions successful in one Y$_{\textrm{H}}$ trap were also successful in a second identically constructed Y$_{\textrm{H}}$ trap, even though the two traps were tested in different vacuum chambers with independent RF and DC voltage sources -- only lasers were shared between the two setups.  These solutions have been successfully used without modification for over six months.

\begin{figure}
\resizebox{1\textwidth}{!}{%
  \includegraphics{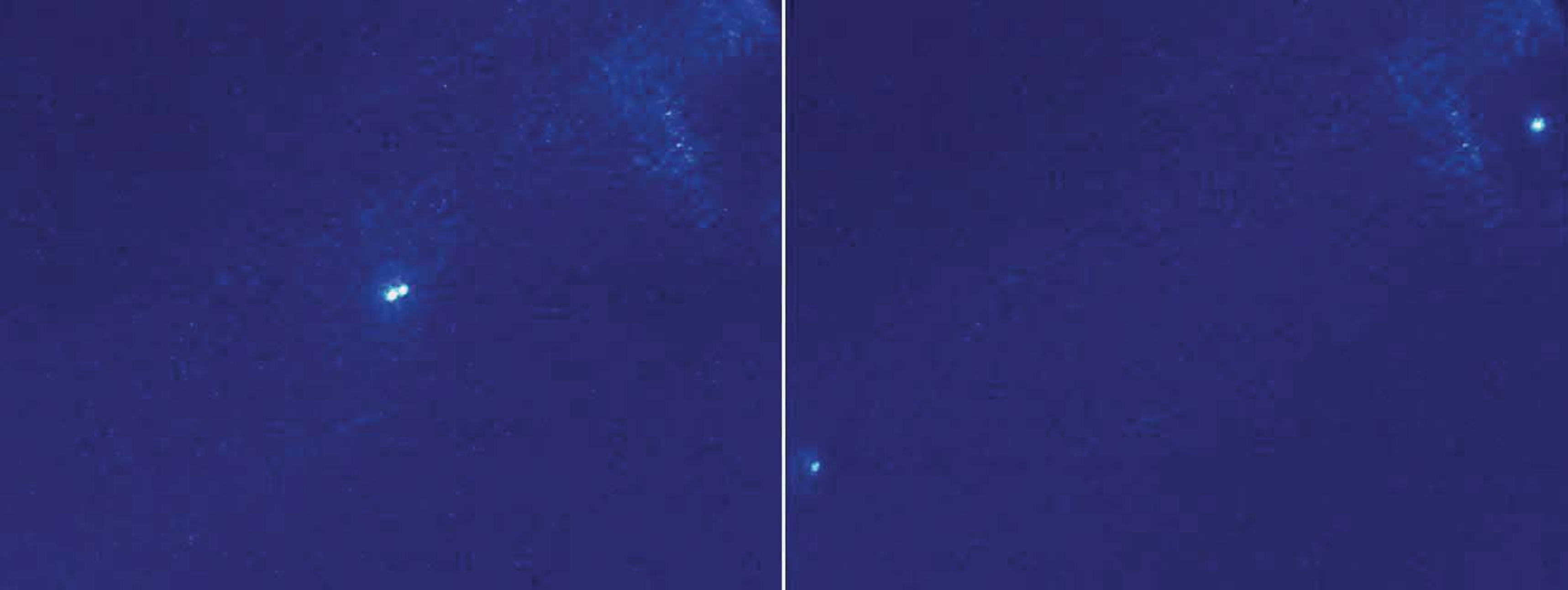}}
\caption{Left: Ions within the same harmonic well separated by approximately 6 microns.  Right: Ions separated by 4 electrodes (approximately 370 microns).  The splitting and recombining of two ions was explicitly observed hundreds of times without error.}
\label{fig:splitting}       
\end{figure}

Finally, complex shuttling routines were implemented for multiple ions in several locations on a given trap.  For instance, three ions were consecutively loaded, independently shuttled into each arm of the junction, and Doppler cooled in a triangular configuration for over an hour without loss.  Linear ion chains were also split and recombined as seen in Fig.~\ref{fig:splitting}.  By performing independent junction shuttling between splitting and recombining, two ions were unequivocally observed to exchange position in a linear chain.  

\section{Conclusion}
\label{sec:conclusion}

Reliable and repeatable micro-fabrication of complex ion-trapping structures has been demonstrated, indicating that a scalable system of trapped ions for quantum computation and quantum simulation is conceivable.  By further utilizing the device design and integration capabilities of multi-layer fabrication techniques, the robust fabrication of more complex 2D and 3D trap arrays is imminent \cite{schmied:2009}.  For example, multi-level metalization (up to four levels of metal) is being employed to accommodate nested trap electrodes and to minimize electrode cross-talk, thereby enabling fundamentally new trap array concepts.

\section{Acknowledgments}

The authors thank the members of Sandia's Microsystems and Engineering Sciences Application (MESA) facility for their fabrication expertise and Mike Descour for helpful comments on the manuscript.  This work was supported by the Intelligence Advanced Research Projects Activity (IARPA).  
Sandia National Laboratories is a multi-program laboratory managed and operated by Sandia Corporation, a wholly owned subsidiary of Lockheed Martin Corporation, for the U.S. Department of Energy's National Nuclear Security Administration under contract DE-AC04-94AL85000.


\begin{thebibliography}{10}
\bibitem{footnote:centerelectrode}
The center DC electrode is directly underneath the ion and extends
  the entire trapping region. A voltage applied to this electrode is applied to
  all regions of the trap.
\bibitem{footnote:reversecompatibility}
Reverse compatibility of the solutions tested in the Y$_{\textrm{H}}$ trap were
  not tested in the Y$_{\textrm{L}}$ trap as it was no longer installed in a
  test setup.
\end{thebibliography}
\end{document}